\documentclass[aps,12pt,final,oneside,onecolumn,nobibnotes,nofootinbib,%
superscriptaddress,centertags,floatfix,secnumarabic,notitlepage]{revtex4}%

\usepackage{bm}
\usepackage{graphics}
\usepackage{rotating}
\usepackage{epsfig}
\usepackage{amsmath}
\usepackage{amsfonts}
\usepackage{varioref}
\usepackage[hidelinks]{hyperref}

\begin{document}

\title{Relativistic corrections to
paired production of charmonium and bottomonium 
in decays of the Higgs boson}

\author{\firstname{R.~N.}~\surname{Faustov}}
\affiliation{Institute of Cybernetics and Informatics in Education, FRC CSC RAS, Moscow, Russia}
\author{\firstname{A.~P.}~\surname{Martynenko}\footnote{{Talk presented at the International Conference on Quantum Field 
Theory, High-Energy Physics, and Cosmology, BLTP JINR, 17-22 July 2022}}}
\affiliation{Samara University, Samara, Russia}
\author{\firstname{F.~A.}~\surname{Martynenko}}
\affiliation{Samara University, Samara, Russia}

\begin{abstract}
\small
Rare decay process of the Higgs boson into a pair of $J/\Psi$ and $\Upsilon$ particles 
is studied within perturbative Standard Model and relativistic quark model.
The relativistic corrections connected with the relative motion of quarks are calculated in the production 
amplitude and the wave functions of the bound states.
Numerical values of the decay widths of the Higgs boson are obtained, 
which can be used for comparison with experimental data.
\end{abstract}



\maketitle

\section{Introduction}

After the discovery of the Higgs boson \cite{atlas,cms}, a period of detailed study of the processes connected 
with this particle was begun. We can say that the study of the Higgs sector has become one of the most important areas 
in particle physics \cite{cms1}.
The discovery of the Higgs boson confirmed the electroweak mechanism of symmetry breaking, but the nature 
of this particle remains to be explored. To do this, it is necessary to create particle colliders, 
on which Higgs bosons could be produced in significant quantities, which would make it possible to proceed 
to a precise study of the parameters of the Higgs sector. It is possible that interaction processes involving 
the Higgs boson can provide a transition to a New Physics that lies beyond the Standard Model.

Among the parameters of the Higgs sector, the coupling constants of the Higgs boson with various bosons g(HZZ), 
g(HWW), leptons $g(H\tau\tau)$, $g(H\mu\mu)$ and quarks g(Hcc), g(Hbb) stand out. They determine the decay processes of the 
Higgs boson into various particles \cite{pdg}. Due to its large mass and the presence of coupling constants with different 
particles, the Higgs boson has numerous decay channels. The decay channel of the Higgs boson into a pair 
of heavy quarks is interesting because it creates the possibility of the production of bound states 
of heavy quarks. Thus, rare exclusive decay processes of the Higgs boson into a pair of charmonium or bottomonium 
are of obvious interest both for studying the decay mechanisms and for studying the properties 
of bound states of quarks.

The CMS collaboration began the search for rare Higgs decays into a pair of heavy vector quarkonia in 2019 \cite{cms2}.
The results of new upper limits on the branching fractions are obtained in \cite{cms3}:
\begin{equation}
B(H\to J/\Psi,J/\Psi)<3.8\cdot 10^{-4},
\end{equation}
\begin{equation}
B(H\to \Upsilon(1S),\Upsilon(1S))<1.7\cdot 10^{-3}.
\end{equation}

Theoretical studies of the production of a pair of heavy quarkonia in the decays of the Higgs boson began 
about 40 years ago in \cite{bander,keung}. In these papers, theoretical formulas were obtained for estimating 
the decay widths in the nonrelativistic approximation for some decay mechanisms. Relatively more recent theoretical 
studies of these processes have been carried out in \cite{luchinsky,apm2021,apm2022,neng}.

In this work we continue the study of relativistic effects in the exclusive 
paired charmonium and bottomonium production in the Higgs boson decay which was begun in \cite{apm2021,apm2022}
for $B_c$ mesons.
Our calculation of the decay widths
is performed on the basis of relativistic quark model used previously for the investigation
of relativistic corrections in different reactions in \cite{apm1,apm2,apm3}.
Despite the rare nature of the Higgs boson decays being studied, one can hope that such processes can be 
investigated at the new Higgs boson factories.
The aim of the present work is to give new analysis of pair quarkonium production in Higgs boson decays in the 
Standard Model:

1. In comparison with previous works \cite{luchinsky,neng} we consider different production mechanisms 
(quark-gluon, quark-photon, photon-photon and Z-boson mechanisms) of the pair charmonium and
bottomonium production.

2. We take into account relativistic corrections connected with the relative motion
of heavy quarks both in the pair production amplitude itself and in the wave functions 
of the bound states.

As a result, new numerical estimation for the Higgs boson decay rates is obtained.
An experimental study of these rare decays of the Higgs boson at future high-luminosity accelerators could 
be useful in testing the Standard Model with greater accuracy.

\section{General formalism}

\begin{figure}[htbp]
\centering
\includegraphics[scale=0.3]{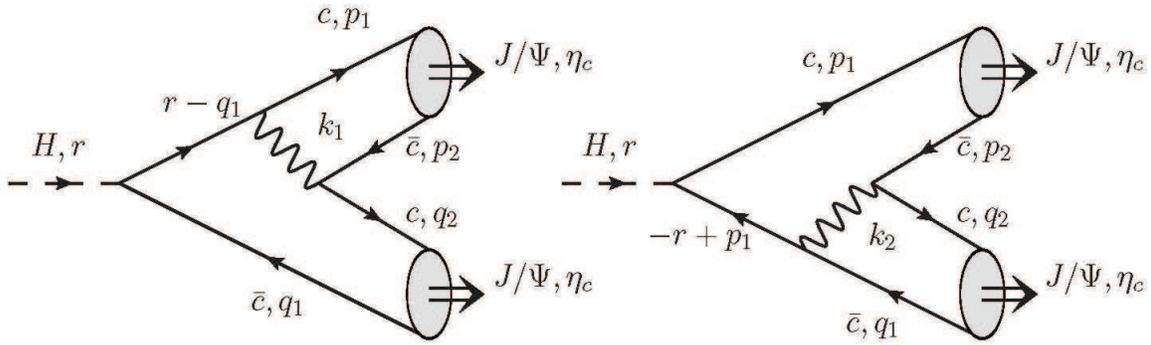}
\caption{Quark-gluon mechanism of the pair charmonium production
in the Higgs boson decay.
Dashed line shows the Higgs boson and wavy line corresponds to the gluon.}
\label{pic1}
\end{figure}

\begin{figure}[htbp]
\centering
\includegraphics[scale=0.3]{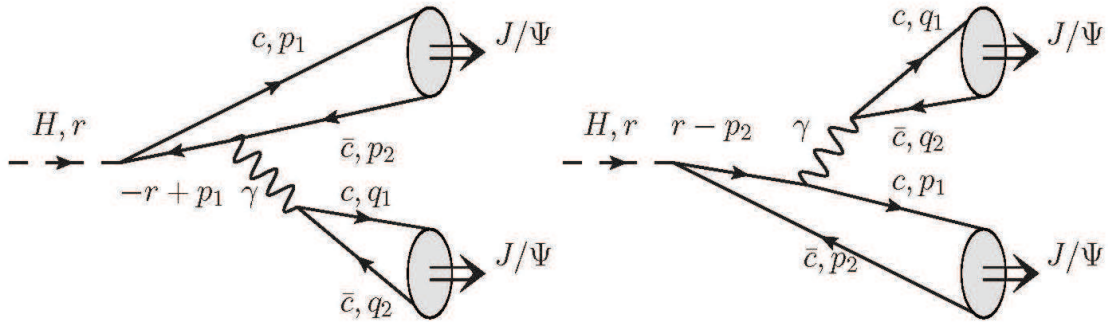}
\caption{Quark-photon mechanism of the pair charmonium production
in the Higgs boson decay.
Dashed line shows the Higgs boson and wavy line corresponds to the photon.}
\label{pic2}
\end{figure}

\begin{figure}[htbp]
\centering
\includegraphics[scale=0.3]{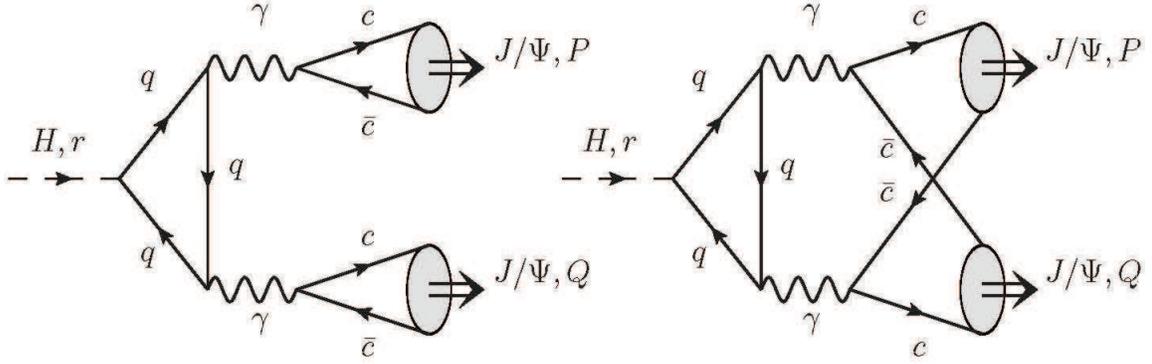}
\caption{Quark loop mechanism of the pair charmonium production
in the Higgs boson decay.
Dashed line shows the Higgs boson and wavy line corresponds to the photon.}
\label{pic3}
\end{figure}

\begin{figure}[htbp]
\centering
\includegraphics[scale=0.3]{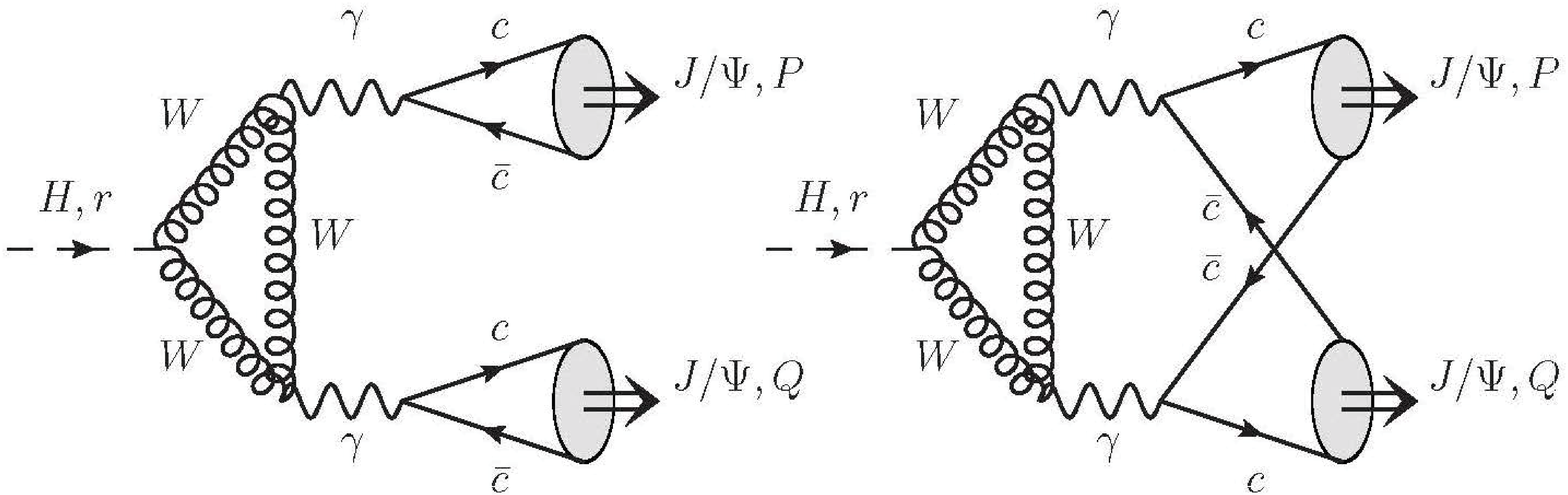}
\includegraphics[scale=0.3]{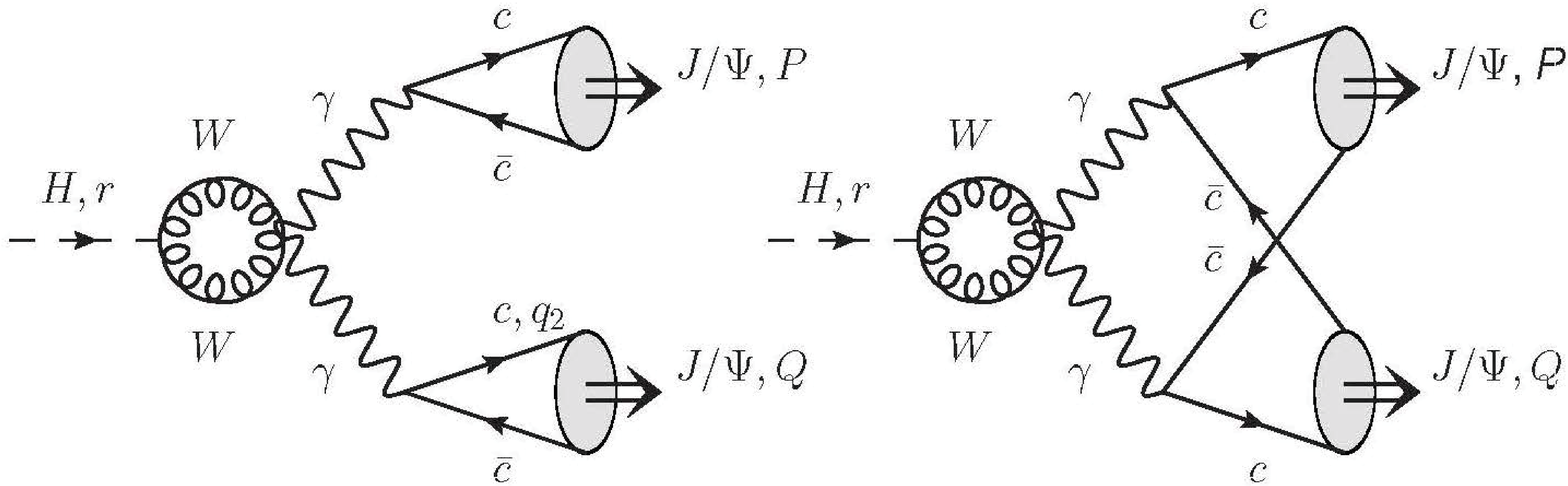}
\caption{W-boson loop mechanism of the pair charmonium production
in the Higgs boson decay.
Dashed line shows the Higgs boson and wavy line corresponds to the photon.}
\label{pic4}
\end{figure}

\begin{figure}[htbp]
\centering
\includegraphics[scale=0.3]{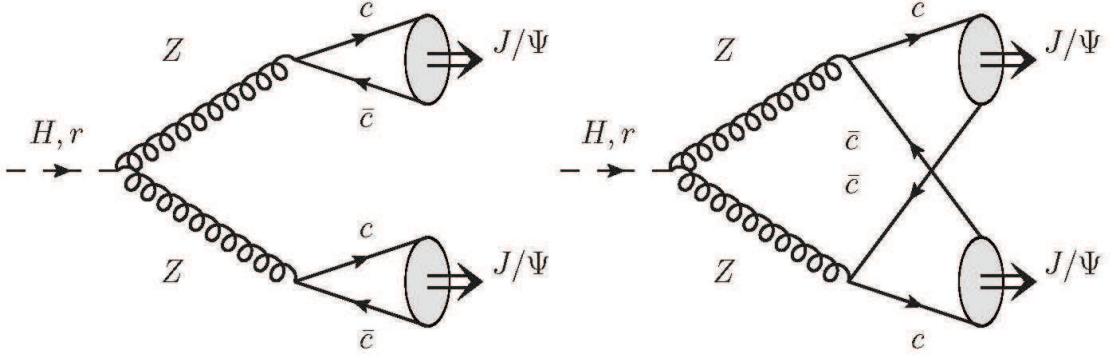}
\caption{Z-boson mechanism of the pair charmonium production
in the Higgs boson decay.
Dashed line shows the Higgs boson and wavy line corresponds to the Z-boson.}
\label{pic5}
\end{figure}

From the very beginning, it should be emphasized that there are several groups of amplitudes for the decay 
of the Higgs boson with the formation of a pair of charmoniums $J/\Psi$ (bottomonium $\Upsilon$), which contribute 
the same order of magnitude to the decay width. We study the amplitudes which are shown in Fig.~\ref{pic1}-Fig.~\ref{pic5} 
and representing different decay mechanisms with pair vector meson production: 
quark-gluon, quark-photon, quark-loop, W-boson loop and Z-boson.
The first group includes quark-gluon amplitudes shown in Fig.~\ref{pic1} (quark-gluon mechanism).
The factor determining the order of contributions can be represented as $\alpha_s/M_H^4$, where $M_H$ is
the mass of the Higgs boson. Such a factor can be distinguished from the very beginning due to the structure 
of the interaction vertices and the denominators of the particle propagators.
The second group is formed by quark-photon amplitudes shown in Fig.~\ref{pic2} (quark-photon mechanism).
The order of contribution is determined here 
by the following factor $\alpha/M_H^2 M_{{\cal Q\bar Q}}^2$. The third group is formed by the amplitudes 
in Fig.~\ref{pic3}-Fig.~\ref{pic4} which contain the quark or W-boson loop with two photons 
that create a pair of $J/\Psi$ (or $\Upsilon$) mesons (quark-loop and W-boson loop mechanisms).
The primary common factor in this case takes the form $\alpha^2/M_{{\cal Q\bar Q}}^4$. 
Finally, the last group of amplitudes in Fig.~\ref{pic5} is determined by the interaction of the Higgs boson 
with a pair of Z-bosons (Z-boson mechanism).
To achieve good calculation accuracy, it is necessary to take into account the contribution 
of all amplitudes from these groups in Fig.~\ref{pic1}-Fig.~\ref{pic5}.

Let consider firstly the Higgs boson decay amplitudes shown in Fig.~\ref{pic1}, \ref{pic2}. For pair production 
of quarkonia in the leading order of perturbation theory, it is necessary to obtain at the first stage two 
free quarks and two free antiquarks. Then they can form bound states with some probability at the next stage.
In the quasipotential approach the decay amplitude
can be presented as a convolution of a perturbative production amplitude
of two $c$-quark and $\bar c$-antiquark pairs and the quasipotential
wave functions of the final mesons \cite{apm2021,apm2022}:
\begin{equation}
\label{eq1}
{\cal M}(P,Q)=-i(\sqrt{2}G_F)^{\frac{1}{2}}\frac{2\pi}{3}M_{{\cal{Q\bar Q}}}
\int\frac{d{\bf p}}{(2\pi)^3}\int\frac{d{\bf q}}{(2\pi)^3}\times
\end{equation}
\begin{displaymath}
\times Tr\left\{\Psi^{\cal V}(p,P)\Gamma_1^{\nu}(p,q,P,Q)\Psi^{\cal V}(q,Q)
\gamma_\nu+
\Psi^{\cal V}(q,Q)\Gamma_2^{\nu}(p,q,P,Q)\Psi^{\cal V}(p,P)
\gamma_\nu\right\},
\end{displaymath}
where $M_{{\cal{Q\bar Q}}}$ is the mass of quarkonium, 
Four-momenta $p_1$ and $p_2$ of $c$-quark and $\bar c$-antiquark in the pair
forming the first ${\cal (Q\bar Q)}$ meson, and four-momenta $q_2$ and $q_1$ for
quark and antiquark in the second meson ${\cal (Q\bar Q)}$ are expressed in terms of relative and total 
four momenta as follows:
\begin{equation}
\label{eq2}
p_{1,2}=\frac12 P \pm p,\quad (pP)=0; \qquad q_{1,2}=\frac12 Q \pm q,\quad (qQ)=0,
\end{equation}
A superscript ${\cal P}$ indicates a pseudoscalar meson ${\cal (Q\bar Q)}$, a superscript ${\cal V}$ 
indicates a vector meson ${\cal (Q\bar Q)}$.
The vertex functions $\Gamma_{1,2}$ are presented below in leading order and in Fig.~\ref{pic1}.
Heavy quarks $c$, $b$ and antiquarks $\bar c$, $ \bar b$ are outside 
the mass shell in the intermediate state:
$p_{1,2}^2\not= m^2$, so that $p_1^2-m^2=p_2^2-m^2$, 
which means that there is a symmetrical exit of particles from the mass shell.

In Eq.~\eqref{eq1} we integrate over the relative three-momenta of quarks and antiquarks
in the final state.
The systematic account of all terms depending on the relative quark momenta $p$ and $q$
in the decay amplitude is
important for increasing the accuracy of the calculation.
$p=L_P(0,\mathbf p)$ and $q=L_Q(0,\mathbf q)$
are the relative four-momenta obtained by the Lorentz transformation of four-vectors
$(0,\mathbf p)$ and $(0,\mathbf q)$ to the reference frames moving with the four-momenta $P$ and $Q$.

The relativistic wave functions of the bound quarks accounting for the transformation from the
rest frame to the moving one with four momenta $P$, and $Q$, are
\begin{equation}
\label{eq3}
\Psi(p,P)=\frac{\Psi_0(\mathbf p)}{\bigl[\frac{\epsilon(p)}{m}
\frac{\epsilon(p)+m}{2m}\bigr]}
\left[
\frac{\hat v_1-1}{2}+\hat v_1\frac{\mathbf p^2}{2m(\epsilon(p)+m)}-\frac{\hat p}{2m}
\right] \times \\
\hat\varepsilon(P,S_z)
\end{equation}
\begin{displaymath}
(1+\hat v_1)
\left[
\frac{\hat v_1+1}{2}+\hat v_1\frac{\mathbf p^2}{2m(\epsilon(p)+m)}+\frac{\hat p}{2m}
\right],
\end{displaymath}
\begin{equation}
\label{eq4}
\Psi(q,Q)=\frac{\Psi_0(\mathbf q)}{\bigl[\frac{\epsilon(q)}{m}
\frac{\epsilon(q)+m}{2m}\bigr]}
\left[\frac{\hat v_2-1}{2}+\hat v_2\frac{\mathbf q^2}{2m(\epsilon(q)+m)}+\frac{\hat q}{2m}
\right] \times \\
\hat\varepsilon(Q,S_z)
\end{equation}
\begin{displaymath}
(1+\hat v_2)
\left[
\frac{\hat v_2+1}{2}+\hat v_2\frac{\mathbf q^2}{2m(\epsilon(q)+m)}-\frac{\hat q}{2m}
\right],
\end{displaymath}
where the hat symbol means a contraction of the four--vector with the Dirac gamma matrices;
$v_1=P/M_{{\cal{Q\bar Q}}}$, $v_2=Q/M_{{\cal{Q\bar Q}}}$; $\epsilon(p)=\sqrt{m^2+\mathbf p^2}$, 
$m$ is $c(b)$-quark mass,
and $M_{{\cal{Q\bar Q}}}$ is the mass of charmonium (bottomonium) state.
$\varepsilon^\lambda(P,S_z)$ is the polarization vector of the $J/\Psi(\Upsilon)$ meson.

Expressions \eqref{eq3} and \eqref{eq4} represent complicated functions depending on relative
momenta ${\bf p}$, ${\bf q}$ including the bound state wave function in the rest frame $\Psi_0({\bf p}) $.
The color part of the meson wave function in the amplitudes \eqref{eq3}-\eqref{eq4} is taken as 
$\delta_{ij}/\sqrt{3}$ (color indexes $i, j, k=1, 2, 3$).

The general structure of expressions \eqref{eq3}-\eqref{eq4} allows us to say that they are the product of the wave 
functions of mesons in the rest frame and special projection operators resulting from the transformation 
from the moving reference frame to the reference frame in which the meson is at rest.
Expressions \eqref{eq3}-\eqref{eq4} make it possible to correctly take into account the relativistic 
corrections connected with the relative momenta of quarks in the final states.
It is useful to note that the expression for the projection operators was obtained in the framework 
of nonrelativistic quantum chromodynamics in \cite{bodwin2002} in a slightly different form for the 
case when the quark momenta lie on the mass shell.
The transformation law of the bound state wave functions of quarks, which is used in the derivation 
of equations \eqref{eq3}-\eqref{eq4} , was obtained in the Bethe-Salpeter
approach in \cite{brodsky} and in quasipotential method in \cite{faustov}.

As follows from \eqref{eq3}-\eqref{eq4}, when constructing the decay amplitudes of the Higgs boson, projection 
operators are introduced for quark-antiquark pairs onto the spin states with total spin $S=1$
of the following form:
\begin{equation}
\label{eq5}
\hat\Pi^{\cal V}=[v_2(0)\bar u_1(0)]_{S=1}=\hat\varepsilon\frac{1+\gamma^0}{2\sqrt{2}}.
\end{equation}

Total amplitude of the Higgs boson decay to paired vector quarkonium in the case of quark-gluon mechanism
in the leading order in strong coupling constant $\alpha_s$ can be presented in the form:
\begin{equation}
\label{eq6}
{\mathcal M}=\frac{4\pi}{3}M_{{\cal{Q\bar Q}}}\alpha_s \Gamma_Q
\int\!\frac{d\mathbf{p}}{(2\pi)^3}\int\!\frac{d\mathbf{q}}{(2\pi)^3}
\mathrm{Tr}\bigl\{\mathcal T_{12}+\mathcal T_{34}\bigr\},
\end{equation}
\begin{equation}
\label{eq7}
{\mathcal T_{12}}=\Psi^{\cal V}(p,P)
\left[\frac{\hat p_1-\hat r+m}{(r-p_1)^2-m^2}\,\gamma_\mu+\gamma_\mu\,
\frac{\hat r-\hat q_1+m}{(r-q_1)^2-m^2}\right]D^{\mu\nu}(k_2)
\Psi^{\cal V}(q,Q)\gamma_\nu,
\end{equation}
\begin{equation}
\label{eq8}
{\mathcal T_{34}}=\Psi^{\cal V}(q,Q)
\left[\frac{\hat p_2-\hat r+m}{(r-p_2)^2-m^2}\,\gamma_\mu+
\gamma_\mu\frac{\hat r-\hat q_2+m}{(r-q_2)^2-m^2}\,\right]D^{\mu\nu}(k_1)
\Psi^{\cal V}(p,P)
\gamma_\nu,
\end{equation}
where $\alpha_{s}=\alpha_s\left(\frac{M_H^2}{4\Lambda^2}\right)$,
$\Gamma_Q=m(\sqrt{2} G_F)^{\frac{1}{2}}$.

Four-momentum of Higgs boson squared $r^2=M_H^2=(P+Q)^2=2M_{{\cal{Q\bar Q}}}^2+2PQ$,
the gluon four-momenta are $k_1=p_1+q_1$, $k_2=p_2+q_2$.
Relative momenta $p$, $q$ of heavy quarks enter in the gluon propagators $D_{\mu\nu}(k_{1,2})$
and quark propagators as well as in relativistic wave functions~\eqref{eq3} and \eqref{eq4}.
Accounting for the small ratio of relative quark momenta $p$ and $q$ to the mass of the Higgs boson $M_H$, 
we can simplify the inverse denominators of quark and gluon propagators as follows:
\begin{equation}
\label{eq9}
\frac{1}{(p_1+q_1)^2}\approx \frac{1}{(p_2+q_2)^2}=\frac{4}{M_H^2},
\end{equation}
\begin{equation}
\label{eq10}
\frac{1}{(r-q_1)^2-m_1^2}=\frac{1}{(-r-p_1)^2-m_1^2}=\frac{1}{(r-p_2)^2-m_1^2}=\frac{1}{(-r-q_2)^2-m_1^2}
=\frac{2}{M_H^2}.
\end{equation}
In \eqref{eq9}-\eqref{eq10} we completely neglect corrections of the form $|{\bf p}|/M_H$, $|{\bf q}|/M_H$. 
At the same time, we keep in the amplitudes \eqref{eq7}, \eqref{eq8} the second-order correction 
for small ratios $|{\bf p}|/m$, $|{\bf q}|/m$ relative to the leading order result.
Calculating the trace in obtained expression
in the package FORM \cite{form}, we find relativistic amplitudes of the paired meson production
in the form:
\begin{equation}
\label{eq12}
{\cal M_{VV}}^{(1)}=\frac{256\pi}{3M_H^4}(\sqrt{2}G_F)^{\frac{1}{2}}m M_{{\cal{Q\bar Q}}} \alpha_s
\varepsilon_1^{\lambda}\varepsilon_2^{\sigma}
F_{1,{\cal VV}}^{\lambda\sigma}|\tilde\Psi_{\cal V}(0)|^2,
\end{equation}
where $\varepsilon_1^\lambda$, $\varepsilon_2^\sigma$ are the polarization vectors of spin 1 
mesons. The superscript in amplitude designation and subscript in tensor function $F_{{\cal VV}}$ designation
denotes the contribution of the quark-gluon mechanism in Fig.~\ref{pic1}.

The contribution of the amplitudes in Fig.~\ref{pic2} must also be taken into account, since these 
amplitudes have the same order as the previous ones, despite the replacement $\alpha_s\to \alpha$.
This is due to the presence in the denominator of the mass of the meson instead of the mass 
of the Higgs boson \cite{bodwin2014}.
The expression for the production amplitudes of the pair $J/\Psi$ has a similar structure with slight 
changes in the common factors:
\begin{equation}
\label{eq13}
{\cal M_{VV}}^{(2)}=\frac{288\pi}{M_H^2M_{\cal{Q\bar Q}}}(\sqrt{2}G_F)^{\frac{1}{2}}m e_Q^2\alpha
\varepsilon_1^{\lambda}\varepsilon_2^{\sigma}
F_{2,{\cal VV}}^{\lambda\sigma}|\tilde\Psi_{\cal V}(0)|^2.
\end{equation}

The tensor corresponding to the quark or W-boson loops in Fig.~\ref{pic3}-Fig.~\ref{pic4}
has the structure (the subscript denotes the contribution of the quark or bosonic loop):
\begin{equation}
\label{eq13d}
T^{\mu\nu}_{Q,W}=A_{Q,W}(t)(g^{\mu\nu}(v_1v_2)-v_1^\nu v_2^\mu)+B_{Q,W}(t)[v_2^\mu-v_1^\mu (v_1v_2)][v_1^\nu-v_2^\nu(v_1v_2)],
\end{equation}
where  $t=\frac{M_h^2}{4m_Q^2}$ or $t=\frac{M_h^2}{4m_W^2}$. The structure functions $A_{Q,W}(t)$, $B_{Q,W}(t)$ can be
obtained using an explicit expression for a loop integrals (see Appendix A,B).

The amplitudes in Fig.~\ref{pic3}-\ref{pic5} with quark, W-boson loops and $ZZ$ in
an intermediate state contain the contributions of direct and crossed diagrams. 
In this case, direct diagrams, in which virtual photons or Z-bosons give vector quarkonia in the final 
state, are dominant in terms of the mass factor. 
However, the structure of the numerators of the direct and cross amplitudes is different, which can 
eventually lead to numerically close contributions.
In what follows, we present the expressions 
for these amplitudes only in the leading order:
\begin{equation}
\label{eq13a}
{\cal M_{VV}}^{(3)}=\frac{2052\pi^2}{m_Q M_{\cal{Q\bar Q}}}(\sqrt{2}G_F)^{\frac{1}{2}}e_q^2 e_Q^2\alpha^2
\varepsilon_1^{\lambda}\varepsilon_2^{\sigma}
F_{3,{\cal VV}}^{\lambda\sigma}|\tilde\Psi_{\cal V}(0)|^2,
\end{equation}
\begin{equation}
\label{eq13b}
{\cal M_{VV}}^{(4)}=\frac{48\pi^2 M_Z M_W}{M^4_{\cal{Q\bar Q}}}(\sqrt{2}G_F)^{\frac{1}{2}} e_q^2\alpha^2\cos\theta_W
\varepsilon_1^{\lambda}\varepsilon_2^{\sigma}
F_{4,{\cal VV}}^{\lambda\sigma}|\tilde\Psi_{\cal V}(0)|^2,
\end{equation}
\begin{equation}
\label{eq13c}
{\cal M_{VV}}^{(5)}=\frac{48\pi\alpha}{M_Z^2\sin^2 2\theta_W}(\sqrt{2}G_F)^{\frac{1}{2}} 
\varepsilon_1^{\lambda}\varepsilon_2^{\sigma}
F_{5,{\cal VV}}^{\lambda\sigma}|\tilde\Psi_{\cal V}(0)|^2,
\end{equation}
where $m$ is the mass of heavy quark c, b, produced in the vertex of Higgs boson decay, $m_Q$ is
the heavy quark mass in quark loop, $e_q$ is the charge (in units e) of heavy quark (c or b) entering
in final mesons, $e_Q$ is the charge (in units e) of heavy quark (c or b) in the quark loop.

The tensor functions in amplitudes \eqref{eq12}-\eqref{eq13c} have the following form:
\begin{equation}
\label{eq15}
F^{\alpha\beta}_{i,{\cal VV}}=g^{(i)}_1 v_1^\alpha v_2^\beta+g^{(i)}_2g^{\alpha\beta},~~~
g^{(1)}_1=-2+\frac{2}{9}\omega_{1}^2,
\end{equation}
\begin{displaymath}
g^{(1)}_2=-1-2r_2+r_1^2+\frac{4}{3}r_2\omega_{1}+\frac{1}{9}\omega_{1}^2+\frac{2}{3}r_2\omega_1^2
-\frac{1}{9}r_1^2\omega_1^2,
\end{displaymath}
\begin{equation}
\label{eq16}
g^{(2)}_1=4-\frac{4}{9}\omega_{1}^2,~
g^{(2)}_2=2+4r_2-2r_1^2-\frac{8}{3}r_2\omega_1-\frac{2}{9}\omega_1^2-\frac{4}{3}r\omega_1^2+\frac{2}{9}r_1^2\omega_{1}^2,
\end{equation}
\begin{equation}
\label{eq16a}
g^{(3)}_{1,Q}= -A_Q(t)(1 + \frac{2}{3}\omega_1 + \frac{1}{9}\omega_1^2)+
B_Q(t)(1 + \frac{2}{3}\omega_1 + \frac{1}{9}\omega_1^2),~
\end{equation}
\begin{displaymath}
g^{(3)}_{2,Q} =A_Q(t)\left(-1 - \frac{2}{3}\omega_1 - \frac{1}{9}\omega_1^2 +\frac{1}{2}r_1^2+\frac{1}{3}\omega_1r_1^2+
\frac{1}{18}\omega_1^2r_1^2\right),
\end{displaymath}
\begin{equation}
g^{(4)}_1=-A_W(t)\left(1+\frac{2}{3}\omega_1+\frac{1}{9}\omega_1^2\right)+B_W(t)\left(1+\frac{2}{3}\omega_1+
\frac{1}{9}\omega_1^2\right),
\end{equation}
\begin{equation}
\label{eq16b}
g^{(4)}_2=A_W(t)\left(-1-\frac{2}{3}\omega_1-\frac{1}{9}\omega_1^2+\frac{1}{2}r_1^2+\frac{1}{3}\omega_1r_1^2+
\frac{1}{18}\omega_1^2r_1^2\right),~
\end{equation}
\begin{equation}
\label{eq16c}
g^{(5)}_{2}=\left(1+\frac{1}{3}\omega_1\right)^2
\left(\frac{1}{2}-a_z\right)^2-\frac{M_z^4}{3\left(\frac{M_h^2}{4}-M_Z^2
\right)^2}\Bigl(-\frac{1}{4}-\frac{1}{6}\omega_1-\frac{1}{36}\omega_1^2+
\end{equation}
\begin{displaymath}
+\frac{1}{2}a_z+\frac{1}{3}\omega_1a_z+\frac{1}{18}\omega_1^2a_z-\frac{1}{2}a_z^2-\frac{1}{3}\omega_1a_z^2-\frac{1}{18}
\omega_1^2a_z^2\Bigr),~~~g^{(5)}_{1}=0,
\end{displaymath}
where the parameter $r_1=\frac{M_H}{M_{{\cal Q\bar Q}}}$, $r_2=\frac{m}{M_{{\cal Q\bar Q}}}$,
$a_z=2|e_Q|\sin^2\theta_W$.

The decay widths of the Higgs boson into a pair of vector quarkonia states are 
determined by the following expressions (see also \cite{apm2021,apm2022}):
\begin{equation}
\label{eq18}
\Gamma_{\cal{VV}}=\frac{2^{14}\sqrt{2}\pi\alpha_s^2 m^2G_F  |\tilde\Psi_{\cal V}(0)|^4 
\sqrt{\frac{r_1^2}{4}-1}}{9M_H^5 r_1^5}\sum_{\lambda,\sigma}
\vert\varepsilon_1^\lambda\varepsilon_2^\sigma F^{\lambda\sigma}_{{\cal VV}}\vert^2,
\end{equation}
\begin{equation}
\label{eq18a}
F^{\lambda\sigma}_{{\cal VV}}=\Bigl[g^{(1)}_1+\frac{9}{16}r_1^2\frac{e_q^2\alpha}{\alpha_s}g^{(2)}_1+
\sum_Q \frac{27\pi}{8}r_1^4\frac{e_Q^2e_q^2\alpha^2m^2_Q}{\alpha_s mM_{{\cal Q\bar Q}}}g^{(3)}_{1,Q}+
\frac{9\pi e_q^2\alpha^2 r_1^4 M_Z M_W}{64\alpha_s m M_{\cal Q\bar Q}}g^{(4)}_1+
\end{equation}
\begin{displaymath}
\frac{9M_H^4\alpha}{16 M_Z^2m M_{\cal Q\bar Q}\alpha_s}\frac{\left(\frac{1}{2}-2|e_q|\sin^2\theta_W\right)^2}{
\sin^22\theta_W}g^{(5)}_{1}
\Bigr]v_1^\sigma v_2^\lambda+
\Bigl[g^{(1)}_2+\frac{9}{16}r_1^2\frac{e_q^2\alpha}{\alpha_s}g^{(2)}_2+
\sum_Q\frac{27\pi}{8}r_1^4\frac{e_q^2e_Q^2\alpha^2 m^2_Q}{\alpha_s m M_{{\cal Q\bar Q}}}g^{(3)}_{2,Q}+
\end{displaymath}
\begin{displaymath}
\frac{9\pi e_q^2\alpha^2 r_1^4 M_Z M_W}{64\alpha_s m M_{\cal Q\bar Q}}\cos\theta_W g^{(4)}_2+
\frac{9M_H^4\alpha}{16 M_Z^2m M_{\cal Q\bar Q}\alpha_s}\frac{1}{
\sin^22\theta_W}g^{(5)}_{2}\Bigr]g^{\lambda\sigma}.
\end{displaymath}
We found it convenient to separate in square brackets in \eqref{eq18a} the coefficients denoting the relative 
contribution of different decay mechanisms with respect to the quark-gluon mechanism in Fig.~\ref{pic1}.
The common factor in \eqref{eq18} corresponds to the amplitudes of the quark-gluon decay mechanism.

\begin{table}[htbp]
\caption{Numerical values of the relativistic parameters~\eqref{eq18}.}
\bigskip
\label{tb1}
\begin{ruledtabular}
\begin{tabular}{|c|c|c|c|c|c|}
$n^{2S+1}L_J$ & Meson &$M^{exp}_{{\cal Q\bar Q}}$, GeV &$\tilde R(0),~~~GeV^{3/2}$ & $\omega_{1}$ & $\omega_{2}$ \\ \hline
$1^1S_0$    & $\eta_c$     &  2.9839  & 0.92 &  0.20    &   0.0087      \\    \hline
$1^3S_1$    & $J/\Psi$     &  3.0969  & 0.81 &  0.20    &   0.0078      \\    \hline
$1^1S_0$    & $\eta_b$     &  9.3987  & 1.95 &  0.05    &   0.0044      \\    \hline
$1^3S_1$    & $\Upsilon$   &  9.4603  & 1.88 &  0.05    &   0.0044      \\    \hline
\end{tabular}
\end{ruledtabular}
\end{table}

General expression for the decay rate \eqref{eq18} contains numerous parameters.
One part of the parameters, such as quark masses, the masses of mesons are determined 
within the framework of quark models as a result of calculating the observed quantities. 
The parameters of quark models are found from the condition of the best agreement with 
experimental data \cite{pdg}.
Another part of the relativistic parameters can also be found in the quark model as 
a result of calculating integrals with wave functions of quark bound states in the 
momentum representation.
Thus, we can say that the approach to calculation based on the relativistic quark model is closed, since 
it allows calculating all the necessary parameters without resorting to additional hypotheses.

The amplitudes of the pair production of vector charmonium and bottomonium in the decay of the Higgs boson 
are expressed in terms of function $F_{VV}$,  
that is presented in the form of an expansion in ${|\bf p}|/m$,
${|\bf q}|/m$ up to terms of the second order.
As a result of algebraic transformations, it turns out to be convenient to express 
relativistic corrections in terms of relativistic parameters
$\omega_{n}$. In the case of S-states $\omega_{n}$ are determined by 
the momentum integrals $I_{n}$ in the form \cite{apm4,apm5}:
\begin{equation}
\label{eq19}
I_{n}^{P,V}=\int_0^\infty p^2R^{P,V}(p)\frac{(\epsilon(p)+m)}
{2\epsilon(p)}
\left(\frac{\epsilon(p)-m}{\epsilon(p)+m}\right)^n dp,
\end{equation}
\begin{equation}
\label{eq20}
\tilde R(0)=\frac{\sqrt{2}}{\sqrt{\pi}}\int_0^\infty \frac{(\epsilon(p)+m)}
{2\epsilon(p)}p^2R(p)dp,~~
\omega^{P,V}_{1}=\frac{I^{P,V}_{1}}{I^{P,V}_{0}},~~
\omega^{P,V}_{2}=\frac{I^{P,V}_{2}}{I^{P,V}_{0}},
\end{equation}
where superscripts $P,V$ denote pseudoscalar and vector states.

\begin{table}[htbp]
\caption{Numerical results for the decay widths in the nonrelativistic approximation
and with the account for relativistic corrections.}
\bigskip
\label{tb2}
\begin{ruledtabular}
\begin{tabular}{|c|c|c|}
Final state & Nonrelativistic decay width   & Relativistic decay width  \\
$(Q\bar Q)(Q\bar Q)$    & $\Gamma_{nr}$    in GeV  &$\Gamma_{rel}$  in GeV \\  \hline
$J/\Psi$ +$J/\Psi$      & $3.29 \cdot 10^{-12}$ &  $0.69 \cdot 10^{-12}$ \\   
$1^3S_1+1^3S_1$         &          &           \\  \hline
$\Upsilon$ +$\Upsilon$  & $0.63 \cdot 10^{-12}$ & $0.74\cdot 10^{-12}$ \\   
$1^3S_1+1^3S_1$         &          &           \\  \hline
\end{tabular}
\end{ruledtabular}
\end{table}

Another source of relativistic corrections is related with the Hamiltonian of the heavy quark 
bound states which allows to calculate 
the bound state wave functions of pseudoscalar and vector mesons (S-states).
The exact form of the bound state wave functions $\Psi_0({\bf q})$
is important to obtain more reliable predictions for the decay widths.
In the nonrelativistic approximation the Higgs boson decay width with a production of a pair 
of quarkonium contains the fourth power of the nonrelativistic wave function at the origin 
for S-states. 
The value of the decay width is very sensitive 
to small changes of $\tilde R(0)$. In the nonrelativistic QCD there 
exists corresponding problem of determining the magnitude of the
color-singlet matrix elements \cite{bbl}. To account for relativistic corrections to the meson 
wave functions we describe the dynamics of heavy quarks by the QCD generalization of
the standard Breit Hamiltonian in the center-of-mass reference frame 
\cite{repko1,pot1,godfrey,lucha1995,rqm2,rqm3,repko2}.

\begin{table}[htbp]
\caption{Numerical values of parameters A and B for W-boson and quark loop mechanisms.}
\bigskip 
\label{tb22}
\begin{ruledtabular}
\begin{tabular}{|c|c|c|}
Parameter  & $H\to J/\psi\:J/\psi$ &$H\to \Upsilon\:\Upsilon$    \\ \hline
$A_c$  & $-2.74\cdot 10^{-4}+1.07\cdot 10^{-4}i$ &  $-3.39\cdot 10^{-3}+0.99\cdot 10^{-3}i$  \\ \hline
$A_b$  & $-5.68\cdot 10^{-5}+7.87\cdot 10^{-4}i$ &  $-1.53\cdot 10^{-3}+0.74\cdot 10^{-3}i$  \\ \hline
$A_t$  & $7.02\cdot 10^{-7}$ &  $6.54\cdot 10^{-6}$  \\ \hline
$A_W$ & $7.94\cdot 10^{-5}$ &   $7.40\cdot 10^{-4}$  \\ \hline
$B_c$  & $0.31\cdot 10^{-4}+0.62\cdot 10^{-10}i$ &  $0.13\cdot 10^{-3}+0.51\cdot 10^{-7}i$  \\ \hline
$B_b$  & $-0.42\cdot 10^{-7}+0.42\cdot 10^{-10}i$ &  $0.37\cdot 10^{-3}+0.35\cdot 10^{-7}i$   \\ \hline
$B_t$ & $0.12\cdot 10^{-14}$ &  $-0.13\cdot 10^{-4}$   \\ \hline
$B_W$ & $1.42\cdot 10^{-11}$ &   $2.76\cdot 10^{-4}$  \\ \hline
\end{tabular}
\end{ruledtabular}
\end{table}

\begin{table}[htbp]
\caption{The contributions of different mechanisms to the Higgs boson decay widths in GeV.}
\bigskip 
\label{tb3}
\begin{ruledtabular}
\begin{tabular}{|c|c|c|}
\multicolumn{3}{|c|}{The contribution accounting for relativistic corrections} \\ \hline
Contribution  & $H\to J/\psi\:J/\psi$ &$H\to \Upsilon\:\Upsilon$    \\ \hline
Fig.\ref{pic1}  & $0.36\cdot 10^{-15}$ &  $0.10\cdot 10^{-12}$  \\ \hline
Fig.\ref{pic2}  & $0.80\cdot 10^{-12}$ &  $0.16\cdot 10^{-12}$   \\ \hline
Fig.\ref{pic3} & $0.70\cdot 10^{-13}$ &  $0.37\cdot 10^{-12}$   \\ \hline
Fig.\ref{pic4} & $0.74\cdot 10^{-13}$ &   $0.68\cdot 10^{-13}$  \\ \hline
Fig.\ref{pic5} & $0.22\cdot 10^{-12}$ &  $1.45\cdot 10^{-12}$   \\ \hline
Total contribution & $0.69\cdot 10^{-12}$   &  $0.74\cdot 10^{-12}$    \\ \hline
\end{tabular}
\end{ruledtabular}
\end{table}

Using the effective Hamiltonian of quarkonia constructed on the basis of perturbative quantum chromodynamics
with allowance for nonperturbative effects, a model of heavy quarkonium for S-states is obtained. Its main 
elements are presented in our previous works \cite{apm2022,apm1,apm2,apm4,apm5}. It was used to calculate 
those relativistic parameters that 
are included in the decay width of the Higgs boson \eqref{eq18}. The numerical values 
of  the parameters $\omega_1$, $\omega_2$ are presented in the Table~\ref{tb1}. The parameter $\omega_{2}$ 
is not included in the decay width, since corrections of the order $O({\bf q}^4)$, $O({\bf p}^4)$ connected with it, 
are omitted.

The results of numerical calculations of the decay widths and parameters A and B
are presented in Tables~\ref{tb2}, \ref{tb22}, \ref{tb3}. 
Comparing the contributions of different meson pair production mechanisms, it is important 
to emphasize that their relative magnitude in the total result is determined by such important parameters 
as $\alpha$ and $\alpha_s$ and the particle mass ratio.
So, for example, in the case of the 
production of a pair of vector mesons, the decay amplitude has two contributions from the gluon (Fig.~\ref{pic1}) 
and photon (Fig.~\ref{pic2}) diagrams. On the one hand, the contribution of photon amplitudes is proportional to $\alpha$, 
which leads to its decrease in comparison with the contribution from gluon amplitudes. But on the other hand, 
the photon contribution contains an additional factor $e_Q^2r_1^2$, which leads 
to an increase in the decay widths. 

\section{Numerical results and conclusion}

The study of rare exclusive decay processes of the Higgs boson is an important task, which makes it possible 
to refine the values of the interaction parameters of particles in the Higgs sector. In this paper, we have attempted 
to present a complete study of decay processes $H\to J/\Psi J/\Psi$, $H\to \Upsilon \Upsilon$
in the Standard Model by considering the various decay mechanisms shown 
in Fig.\ref{pic1}-Fig.\ref{pic5}. To improve the calculation accuracy, we take into account 
relativistic corrections, which were not previously considered in \cite{luchinsky,neng}. 
The numerical contribution of different decay mechanisms 
is analyzed and it is shown that in the case of double production of charmonium, the main decay mechanism 
is the quark-photon one in Fig.\ref{pic2}, ZZ-mechanism in Fig.\ref{pic5}
although taking into account other mechanisms is also necessary to achieve high calculation accuracy
(see also \cite{bodwin2014}). For bottomonium pair production 
all mechanisms give close contributions to the width, but the ZZ-mechanism is dominant.
Their sum ultimately determines the full numerical result.
The parameters A and B, which determine the numerical values of the contributions 
of the W-boson loop mechanism and the quark loop mechanism, are obtained in analytical and numerical form. 
They are presented separately in Table~\ref{tb22}.

In our approach, we distinguish two types of relativistic corrections. The corrections of the first type are 
determined by the relative momenta p and q in the production amplitude of two quarks and two antiquarks. 
The corrections of the second type are determined by the transformation law of the meson wave functions, 
which results in expressions \eqref{eq3}-\eqref{eq4}. The $\psi({\bf p})$ wave functions entering into
\eqref{eq3}-\eqref{eq4} in the 
rest frame of the bound state are found from the solution of the Schr\"odinger equation with a potential that includes relativistic corrections. Having an exact expression for the decay amplitude (see the example in the Appendix C),
we carry out a series of transformations with it, extracting second-order corrections in p and q. The model used 
is described in more detail in \cite{apm2022}. In our approach, all arising relativistic parameters $\omega_1$,
$\omega_2$, $\tilde R(0)$ are determined within the framework of the relativistic quark model (see Table~\ref{tb1}).
Accounting for relativistic corrections in this work shows that such contributions lead to 
a significant change in nonrelativistic results. Here it must be emphasized that we call nonrelativistic 
results such results that are obtained at $p=0$, $q=0$ and neglecting relativistic corrections in the quark 
interaction potential. The main parameter that greatly reduces the nonrelativistic results is $\tilde R(0)$, 
which enters the fourth power of the decay width.
Therefore, the difference between the relativistic and nonrelativistic results in Table~\ref{tb2} turns out 
to be more significant in the case of pair production of charmonium.

The paper considers various mechanisms for the production of a pair of vector mesons in the decay of the 
Higgs boson, which are presented in Fig.~\ref{pic1}-Fig.~\ref{pic5}. There are other amplitudes 
for the production of a pair of $J/\Psi$, $\Upsilon$ mesons, 
which are calculated but not included in detail in the work. So, for example, there is a pair production mechanism 
(HH mechanism), when the original Higgs boson turns into a HH pair, which then gives a pair of vector mesons 
in the final state. In a sense, it is similar to the ZZ mechanism, but the amplitude structure is different, 
which results in a contribution to the decay width that is several orders of magnitude smaller than those 
given in Table~\ref{tb1}.
Despite the obvious difference in the amplitudes in Fig.~\ref{pic1} and Fig.~\ref{pic2} in terms of the mass factor
$M^2/M_h^2$, 
we include the amplitudes of Fig.~\ref{pic1} in the consideration, in contrast to work \cite{neng}.
We do not consider amplitudes like those shown in Fig.~\ref{pic3}, but with two gluons. In this case, the direct 
amplitude (Fig.~\ref{pic3} (left)) vanishes due to the color factor, and the cross amplitude 
(Fig.~\ref{pic3} (right)) is suppressed by the mass factor $M^2/M_h^2$.
A distinctive feature of our calculations is that when studying the contributions of the amplitudes 
in Fig.~\ref{pic3}-\ref{pic4}, we take into account two structure functions A and B \eqref{eq13d} in the tensor 
function of the triangular loop. They are calculated analytically and numerically, and the corresponding 
results are presented in Table~\ref{tb3} and Appendixes A and B.
Although the function B initially contains higher powers of the factor $r_4$ (see Appendix A, B), 
which should lead to a decrease in the numerical values for $r_4 \ll 1$, nevertheless, the structure 
of the coefficient functions in \eqref{eq18} is such that the function B cannot be neglected.

In Table~\ref{tb3}, we present separately the numerical values of the contributions from different decay mechanisms 
with an accuracy of two significant figures after the decimal point. First of all, the problem was to obtain 
the main contribution to the decay width. This was the contribution from the quark-photon mechanism in 
Fig.~\ref{pic2} and the ZZ mechanism in Fig.~\ref{pic5}, which are one order of magnitude 
greater than the other contributions in the case of charmonium production. Its value is due to the structure 
of the decay amplitudes, in which small denominators $1/M^2$ appear from the photon propagators
in contrast to other amplitudes in which there is a factor $1/M_H^2$. In addition, 
the numerator in these amplitudes contains amplifying factors in powers of a large parameter $r_1$. In each 
considered decay mechanism, there are radiative corrections $O(\alpha_s)$ that were not taken into account. They represent 
a major source of theoretical uncertainty, which we estimate to be about $30\%$ in $\alpha_s$. Other available 
errors connected with the parameters of the Standard Model and higher order relativistic corrections 
do not exceed $10\%$.
On the whole, our complete numerical estimates of the decay widths agree in order of magnitude 
with the results \cite{neng}.

\acknowledgments
The work is supported by the Foundation for the Advancement 
of Theoretical Physics and Mathematics "BASIS" (grant No. 22-1-1-23-1). 
The authors are grateful to A.~V.~Berezhnoy, I.~N.~Belov, D.~Ebert, V.~O.~Galkin, A.~K.~Likhoded
for useful discussions.

\appendix
\section{The calculation of W-boson loop by dispersion method}
\label{app1} 
In two Appendices A and B we consider the calculation of tensors that determine the contributions 
to the Higgs boson decay width from the two mechanisms shown in Fig.~\ref{pic3} and Fig.~\ref{pic4}
(see also \cite{w1,w2,apm1998,w3,w4,w5,w6,w7,w8,w9,w10}).

The Higgs boson decay amplitude into two charmonium states contains the contribution which is determined
by W-boson loop presented in Fig.~\ref{pic4}.
Generally speaking, along with W-boson loops, it is also necessary to consider other many-numbered loop 
contributions determined by Goldstone bosons and ghosts. But there is a unitary gauge convenient for calculation, 
in which there is only the contribution of the diagrams presented in Fig.~\ref{pic4}.

The tensor corresponding to triangle loop can be written as follows:
\begin{equation}
\label{eqa1}
T^{\mu\nu}_{1,W}=8\pi\alpha M_ZM_W(\sqrt{2}G_F)^{1/2}\cos\theta_W\int\frac{d^4 k}{(2\pi)^4}
\frac{V_{\mu\sigma\rho}(r-k,P,Q-k)V_{\nu\omega\lambda}(Q-k,Q,-k)}{(k^2-M_W^2)((r-k)^2-M_W^2)((Q-k)^2-M_W^2)}\times
\end{equation}
\begin{displaymath}
\left[g^{\rho\omega}-\frac{(Q-k)^\rho(Q-k)^\omega}{M_W^2}\right]
\left[g^{\lambda\alpha}-\frac{(r-k)^\lambda(r-k)^\alpha}{M_W^2}\right]
\left[g^{\sigma\alpha}-\frac{ k^\sigma k^\alpha}{M_W^2}\right]+(\mu\leftrightarrow\nu,~P\leftrightarrow Q).
\end{displaymath}
The tensor $T^{\mu\nu}_{2,W}$ of second Feynman amplitude in Fig.~\ref{pic4} has the similar form. Next, 
we consider the sum of these two amplitudes.

To calculate the tensor $T^{\mu\nu}_W$ of bosonic loop, we use the dispersion method, keeping in mind that 
the tensor that defines this loop has the general structure \eqref{eq13d}. 
Then the structure functions $A_W(t)$, $B_W(t)$ can be obtained by performing 
the following convolutions over the Lorentz indices:
\begin{equation}
\label{eqa2}
A_W(t)=T^{\mu\nu}\frac{1}{2}\left[\frac{g^{\mu\nu}}{v_1v_2}-\frac{v_1^\nu v_2^\mu}{(v_1 v_2)^2-1}\right],~~~
B_W(t)=T^{\mu\nu}\frac{1}{2}\left[\frac{3v_1^\nu v_2^\mu}{((v_1 v_2)^2-1)^2}-
\frac{g^{\mu\nu}}{v_1v_2((v_1 v_2)^2-1)}\right].
\end{equation}

For two structure functions $A_W(t)$, $B_W(t)$ we use the dispersion relation with one subtraction: 
\begin{equation}
\label{eqa3}
A_W(t)=A_W(0)+\frac{t}{\pi}\int_1^\infty\frac{Im A(t') dt'}{t'(t'-t+i0)},~~~
B_W(t)=B_W(0)+\frac{t}{\pi}\int_1^\infty\frac{Im B(t') dt'}{t'(t'-t+i0)}.
\end{equation}

Imaginary parts $Im A(t)$, $Im B(t)$ can be calculated using the Mandelstam-Cutkosky rule \cite{rec,t4}.
For example, the imaginary part of the function $A(t)$ is equal
\begin{equation}
\label{eqa4}
Im A_W=\frac{r_3^2}{512\pi\sqrt{t(t-1)}(2t-r_3^2)(t-r_3^2)^{3/2}}
\Bigl\{4r_3^2(1-t)(t-r_3^2)^{1/2}\Bigl[r_3^2(4+r_3^2-2t)(2t+1)-4(2t+3)\Bigr]
\end{equation}
\begin{displaymath}
+\sqrt{t-1}\Bigl[8r_3^6+48(1-2t)t+r_3^8(1+2t)-4r_3^4(3+t)(3+2t)+16r_3^2(-3+t(9+2t))\Bigr]\times
\end{displaymath}
\begin{displaymath}
\ln\frac{r_3^2-2t+2\sqrt{(t-1)(t-r_3^2)}}{r_3^2-2t-2\sqrt{(t-1)(t-r_3^2)}}\Bigl\},~~~
r_3=\frac{M}{M_W},~~M=M_{{\cal Q\bar Q}}.
\end{displaymath}

This expression accurately takes into account the dependence on the meson mass $M$. Accounting for 
$r_3\ll 1$, we can perform an expansion in $r_3$:
\begin{equation}
\label{eqa5}
Im A_W(t)=\frac{1}{16\pi}\Biggl\{\frac{3r_3^2}{4}\frac{(2t-1)\ln\frac{\sqrt{t}+\sqrt{t-1}}{\sqrt{t}-
\sqrt{t-1}}}{t^2}+
\end{equation}
\begin{displaymath}
\frac{r_3^4}{t^3\sqrt{t-1}}\left[\sqrt{t}(2t^2+t-3)-\sqrt{t-1}(3-3t+2t^2)\ln\frac{\sqrt{t}+\sqrt{t-1}}{\sqrt{t}-
\sqrt{t-1}}
\right]\Biggr\}.
\end{displaymath}

Calculating the dispersion integral with subtraction, we get:
\begin{equation}
\label{eqa6}
A_W(t)=\frac{1}{16\pi^2}r_3^2\tilde A_W(t)=
\frac{1}{16\pi^2}r_3^2\Bigl\{\frac{3}{t^2}\left[t+(2t-1)f(t)^2\right]-5\Bigr\}+
\end{equation}
\begin{displaymath}
\frac{1}{16\pi^2}\frac{2r_3^4}{15 t^4}\Bigl\{90t-15t^2+7t^3+15f(t)\Bigl[
\frac{(2t^2+t-3)}{\sqrt{t-1}t^{7/2}}+(-3+3t-2t^2)f(t)\Bigr]\Bigr\},
\end{displaymath}
\begin{equation}
\label{eqa7}
f(t)=
\begin{cases}
\arcsin\sqrt{t},~~~t\leq 1,\\
\frac{i}{2}\left[\ln\frac{1+\sqrt{1-t^{-1}}}{1-\sqrt{1-t^{-1}}}-i\pi\right],~~~t>1,
\end{cases}
\end{equation}
\begin{equation}
\label{eqa7a}
B_W(t)=\frac{1}{16\pi^2}\frac{r_3^6}{210t^9(1-t)}\Bigl\{t^5(t-1)\Bigl[-t(1470+t(2051+1943t))+210\ln(1-t)
\times
\end{equation}
\begin{displaymath}
(-6_t(-4+t(-3+2t)))+4\ln 2(315+t(735+44t(7+16t)))\Bigr]+105i\sqrt{t^9(1-t)}(3+5t-2t^2+4t^3)\times
\end{displaymath}
\begin{displaymath}
\Bigl[\ln 2\ln\frac{1-2t-2i\sqrt{t(1-t)}}{1-2t+2i\sqrt{t(1-t)}}+
2Li_2\left(t-i\sqrt{t(1-t)}\right)-2Li_2\left(t+i\sqrt{t(1-t)}\right)-
\end{displaymath}
\begin{displaymath}
Li_2\left(2t-2i\sqrt{t(1-t)}\right)+Li_2\left(2t+2i\sqrt{t(1-t)}\right)\Bigr]\Bigr\}.
\end{displaymath}

The subtraction constant $\tilde A_W(0)=7$. Then, in the leading order in $r_3$, 
the function $A_W(t)$ is equal to
\begin{equation}
\label{eqa8}
A_W(t)=\frac{1}{16\pi^2}r_3^2\left[2+\frac{3}{t}+\frac{3}{t^2}(2t-1)f(t)^2\right].
\end{equation}

The limit $M_W\to 0$ within the Goldstone-boson approximation gives: $\tilde A_W|_{M_W\to 0}\to  2$.
The numerical value of $B_W(t)$ is significantly less than \eqref{eqa6} due to the factor $r_3^6$.

\section{The calculation of quark loop by dispersion method}
\label{app2} 
The calculation of the quark loops contribution to structure functions $A_Q(t)$, $B_Q(t)$
can be carried out also by dispersion method. Final expression for imaginary parts of
these functions are the following:
\begin{equation}
\label{eqb1a}
Im A_Q=-\frac{r_4^2}{512\pi\sqrt{t}(2t-r_4^2)(t-r_4^2)^{2}}
\Bigl\{4r_4^2(r_4^2-t)\sqrt{t-1}+
\end{equation}
\begin{displaymath}
\sqrt{t-r_4^2}\Bigl[r_4^4+r_4^2(4-6t)+4t(t-1)\Bigr]
\ln\frac{r_4^2-2t+2\sqrt{(t-1)(t-r_4^2)}}{r_4^2-2t-2\sqrt{(t-1)(t-r_4^2)}}\Bigr\},~~r_4=\frac{M}{m_Q},
\end{displaymath}
\begin{equation}
\label{eqb1}
Im B_Q=\frac{r_4^6}{256\pi t^{3/2}(2t-r_4^2)(t-r_4^2)^{5/2}}
\Bigl\{-4(r_4^2-4t)\sqrt{(t-1)(t-r_4^2)}+
\end{equation}
\begin{displaymath}
\Bigl[r_4^4+4t(1+2t)-2r_4^2(2+3t)\Bigr]
\ln\frac{r_4^2-2t+2\sqrt{(t-1)(t-r_4^2)}}{r_4^2-2t-2\sqrt{(t-1)(t-r_4^2)}}\Bigr\},~~r_4=\frac{M}{m_Q},
\end{displaymath}
where $m_Q$ is the mass of heavy quark in the loop.

In the case of quark loops, we use the dispersion relation without subtractions. 
The functions $A_Q(t)$ and $B_Q(t)$ have the following explicit form after expansion in $r_4^2$:
\begin{equation}
\label{eqb2}
A_Q(t)=
-\frac{1}{16\pi^2}\frac{r_4^2}{32 t^3}\Bigl\{4t^2+4t(t-1)f^2(t)+r_4^2\Bigl[
-2t(t-4)-4\sqrt{t(1-t)}f(t)+2(t-2)f^2(t)\Bigr]\Bigl\},
\end{equation}
\begin{equation}
\label{eqb3}
B_Q(t)=-\frac{1}{16\pi^2}\frac{r_4^6}{2520 t^5}\Bigl\{2t\Bigl[-14t(225+t(45+14t))-r_4^2(6930+t(1365+t(413+216t)))\Bigr]+
\end{equation}
\begin{displaymath}
1260\Bigl[9r_4^2\sqrt{(1-t)t}+4\sqrt{(1-t)t^3}\Bigr]f(t)+630(2t(1+2t)+r_4^2(4+9t))f^2(t)\Bigr\}.
\end{displaymath}

In the leading order in $r_4^2$ the expression \eqref{eqb1} coincides with the known expression obtained
in \cite{w1,w2,apm1998,w3,w4,w5,w6,w7,w8,w9}. Since $B(t)$ contains the factor $r_4$ to a higher degree, 
its contribution is sometimes neglected.

\section{The amplitude of pair vector mesons production via ZZ mechanism}
\label{app3} 
The amplitude of direct production of vector mesons (Fig.~\ref{pic5}(left)) has the form:
\begin{equation}
\label{eqc1}
{\mathcal M}_{dir}=\frac{48\pi\alpha (\sqrt{2}G_F)^{1/2}M_Z^2}{\sin^22\theta_W}
\int\frac{d\mathbf{p}}{(2\pi)^3}\frac{\Psi_0({\bf p})}
{\frac{\epsilon(p)}{m}\frac{(\epsilon(p)+m)}{2m}}
\int\frac{d\mathbf{q}}{(2\pi)^3}\frac{\Psi_0({\bf q})}
{\frac{\epsilon(q)}{m}\frac{(\epsilon(q)+m)}{2m}}D_{\mu\alpha}(P)D_{\nu\alpha}(Q)
\times
\end{equation}
\begin{displaymath}
Tr\Bigl\{[\frac{\hat v_1-1}{2}-\frac{\hat v_1p^2}{2m(\epsilon(p)+m)}-\frac{\hat p}{2m}]
\hat\varepsilon_1(\hat v_1+1)
[\frac{\hat v_1-1}{2}-\frac{\hat v_1p^2}{2m(\epsilon(p)+m)}+\frac{\hat p}{2m}]
\gamma_\mu[\frac{1-\gamma_5}{2}-a_z]
\Bigr\}\times
\end{displaymath}
\begin{displaymath}
Tr\Bigl\{[\frac{\hat v_2-1}{2}-\frac{\hat v_2q^2}{2m(\epsilon(q)+m)}-\frac{\hat q}{2m}]
\hat\varepsilon_2(\hat v_2+1)
[\frac{\hat v_2-1}{2}-\frac{\hat v_2q^2}{2m(\epsilon(q)+m)}+\frac{\hat q}{2m}]
\gamma_\nu[\frac{1-\gamma_5}{2}-a_z]
\Bigr\},
\end{displaymath}
where $\varepsilon_{1,2}$ are the polarization four-vectors describing vector meson states.
$D_{\mu\alpha}(k)$ is the Z-boson propagator.

The amplitude of crossed production of vector mesons (Fig.~\ref{pic5}(right)) has the form:
\begin{equation}
\label{eqc2}
{\mathcal M}_{cr}=\frac{16\pi\alpha (\sqrt{2}G_F)^{1/2}M_Z^2}{\sin^22\theta_W}
\int\frac{d\mathbf{p}}{(2\pi)^3}\frac{\Psi_0({\bf p})}
{\frac{\epsilon(p)}{m}\frac{(\epsilon(p)+m)}{2m}}
\int\frac{d\mathbf{q}}{(2\pi)^3}\frac{\Psi_0({\bf q})}
{\frac{\epsilon(q)}{m}\frac{(\epsilon(q)+m)}{2m}}D_{\mu\alpha}(P)D_{\nu\alpha}(Q)
\times
\end{equation}
\begin{displaymath}
Tr\Bigl\{[\frac{\hat v_1-1}{2}-\frac{\hat v_1p^2}{2m(\epsilon(p)+m)}-\frac{\hat p}{2m}]
\hat\varepsilon_1(\hat v_1+1)
[\frac{\hat v_1-1}{2}-\frac{\hat v_1p^2}{2m(\epsilon(p)+m)}+\frac{\hat p}{2m}]
\gamma_\mu[\frac{1-\gamma_5}{2}-a_z]\times
\end{displaymath}
\begin{displaymath}
[\frac{\hat v_2-1}{2}-\frac{\hat v_2q^2}{2m(\epsilon(q)+m)}-\frac{\hat q}{2m}]
\hat\varepsilon_2(\hat v_2+1)
[\frac{\hat v_2-1}{2}-\frac{\hat v_2q^2}{2m(\epsilon(q)+m)}+\frac{\hat q}{2m}]
\gamma_\nu[\frac{1-\gamma_5}{2}-a_z]
\Bigr\}.
\end{displaymath}
After all the transformations, the total amplitude of the production of a pair 
of vector mesons in the ZZ mechanism takes the form \eqref{eq13c}.

\end{document}